\documentclass{article}
\usepackage{epsfig}
\usepackage{graphicx}
\usepackage{amsfonts}
\usepackage{euscript}
\usepackage[cp1251]{inputenc}
\usepackage[russian]{babel}
\usepackage{amsmath}
\usepackage{amssymb}
\begin{document}

Our examination of a binary star mass distribution
\cite{V-01,m-star} shows that cores of these stars in their stable
state should be composed from plasma consisting of electrons and
neutron-excess nuclei (the same as $^3H,~^4H,~^5H$ and
$~^6He,~^8He,~^{10}He$). These nuclei are $\beta$-active under
"normal" terrestrial conditions. They get stable in central regions
of stars under pressure of plasma electron gas. This pressure
averaged over a plasma cell at plasma density about $10^{25}$
particale/$cm^3$ and temperature about $10^{7}$K \cite{V-01,m-star}
amounts to $10^{6}$GPa. It can be considered as the lowest bond of
the electron pressure acting on nuclei. As the density of electron
gas in the nearest vicinity of a nucleus is much more than the
density averaged over a cell volume, its pressure on nucleus is
substantially higher than the mentioned value. It is impossible to
achieve Such a pressurev at laboratory conditions, as well as to
influence the rate of $\beta$-decay. Let us estimate how much can
the electron gas slow down the decay rate at laboratory conditions.
The most suitable object for this investigation is tritium. It is
subjected to $\beta$-decay with a highest possible energy of an
outgoing electron equal to $E_\beta$=18.6 kev. The ionization energy
of hydrogen is $E_i$=13.6~ev. The energy of this level can be held
by electrons in plasma, which is created by a gas discharge in
tritium gas. If nuclei experiencing $\beta$-decay compose atomic
substance, they emit a full spectrum of electron energy from 0 to
$E_\beta$. If these nuclei compose an electron-nuclear plasma, the
outgoing of decay electrons with energies less than $E_i$ must be
limited by the Pauli principle. The part of a phase volume of this
state is
\begin{equation}
\delta=\left(\frac{E_i}{E_\beta}\right)^{3/2}\approx 10^{-4.5}
\end{equation}
The probability of a electron outgoing with energy close to the
threshold $E_\beta$ is small in comparison to the probability of a
electron outgoing with a small energy $E\ll E_\beta$. It must
several (approximately 3) times increase the effect of plasma
influence and we can expect that the influence of plasma on
$\beta$-decay reaches a value approximately equal to $10^{-4}$. To
have the statistically valid measured result, the total amount N of
decays during the time of measurement must be at the level
\begin{equation}
N>\left(\frac{3}{\delta}\right)^2\approx 10^{8},
\end{equation}
It seems quite realistic to achieve this effect with enough tritium
in the experimental device and a reasonable rate and time of the
measurement.


\begin{thebibliography}{}
\bibitem {V-01}   Vasiliev B.V. - Nuovo Cimento B, 2001, v.116, pp.617-634.
\bibitem {m-star} Vasiliev B.V. - Astro-ph/0512468
\end{thebibliography}
\end{document}